\documentclass[prb,twocolumn,showpacs,preprintnumbers]{revtex4}%
\usepackage{graphicx}
\usepackage{dcolumn}

\usepackage{bm}%
\usepackage{amsmath}%
\usepackage{amsfonts}%
\usepackage{amssymb}

\begin{document}
\title{Edge Channel Transport in InAs/GaSb Topological Insulating Phase}
\author{Kyoichi Suzuki}
\email{suzuki.kyoichi@lab.ntt.co.jp}
\author{Yuichi Harada}
\author{Koji Onomitsu}
\author{Koji Muraki}
\affiliation{NTT Basic Research Laboratories, NTT Corporation,}
\affiliation{3-1 Morinosato-Wakamiya, Atsugi, Kanagawa 243-0198, Japan }
\date{\today}

\begin{abstract}
Transport in InAs/GaSb heterostructures with different InAs layer thicknesses
is studied using a six-terminal Hall bar geometry with a 2-$\mu$m edge channel
length. For a sample with a 12-nm-thick InAs layer, non-local resistance measurements with
various current/voltage contact configurations reveal that the transport is
dominated by edge channels with negligible bulk contribution. Systematic
non-local measurements allow us to extract the resistance of individual edge
channels, revealing sharp resistance fluctuations indicative of inelastic
scattering. Our results show that the InAs/GaSb system can be tailored to have
conducting edge channels while keeping a gap in the bulk region and provide 
a way of studying 2D topological insulators even when quantized transport is absent.

\end{abstract}

\pacs{72.25.Dc, 73.63.Hs, 73.61.Ey, }
\maketitle



\section{INTRODUCTION}

Topological insulators (TIs) have attracted strong interest as a new quantum state of
matter categorized neither as a metal nor an insulator.
\cite{QSHE1-1,QSHE1-2,QSHE1-3,QSHE2-1,QSHE2-2,QSHE3-1,QSHE3-2,HgTe1,HgTe2,HgTe3,HgTe4} 
Due to their topological nature, gapless states are formed at the sample
periphery, i.e., at surfaces and edges in three-dimensional
\cite{QSHE2-1,QSHE2-2,QSHE3-1,QSHE3-2} and two-dimensional (2D) TIs, 
\cite{HgTe1,HgTe2,HgTe3,HgTe4} 
respectively. A unique feature of 2D TIs is
that their edge channels are helical; that is, each edge comprises two
channels carrying the two spin components in opposite directions. In the
absence of a magnetic field or inelastic scattering, transport is protected
from backscattering by time-reversal symmetry. If the sample dimension is
smaller than the inelastic scattering length, the edge transport becomes
dissipationless, which would lead to quantized spin Hall conductance. The
observation of a quantum spin Hall effect (QSHE) in HgTe/HgCdTe quantum wells
\cite{HgTe2,HgTe3} and subsequent non-local resistance measurements
\cite{HgTe4}
have confirmed the formation of a 2D TI phase. However, the experimental
demonstration of a 2D TI can be affected by sample disorder, which limits the
inelastic mean free path \cite{HgTe2,HgTe3,HgTe4,supporting,Glazman} and hence hampers
the observation of a QSHE or quantization of non-local transport. Since the
existence of edge states in a TI stems from its bulk properties, \cite{QSHE2-1,
MurakamiNJP} a more basic question is whether or not the material has the band
structure characteristic of a TI.

Until now, all experimentally confirmed TIs, \cite{QSHE2-1,QSHE2-2,QSHE3-1,QSHE3-2,HgTe2,HgTe3,HgTe4}   
such as Bi$_{1-x}$Sb$_{x}$, Bi$_{2}$Te$_{3}$, TlBiSe$_{2}$, and HgTe, naturally have the band
structure characteristics of TIs with band inversion caused by spin-orbit
interaction. In contrast, the InAs/GaSb hybrid system has been predicted to
offer a new type of TI consisting solely of semiconductors with non-inverted
band structures and additional band tunability via an electric field. \cite{InAs_GaSb} 
For appropriately designed InAs and GaSb layer thicknesses,
the bottom of the InAs conduction band becomes lower in energy than the top of the GaSb valence band. 
Due to band hybridization across the heterointerface and spin-orbit
interaction, a small energy gap opens at the crossing points of the original
bands, \cite{Altarelli,SuzukiPRL,minigap} 
and helical edge channels are expected to be formed. Recently, Knez and
co-workers have investigated InAs/GaSb four-terminal devices. \cite{Du} 
From systematic conductance measurements on samples with various dimensions and
length/width aspect ratios, they were able to identify the contribution of
edge channel transport with conductance close to the expected quantized value.
However, the data indicated significant contributions of bulk transport,
making it unfeasible to obtain more direct information about the edge transport.

In this paper, we report transport measurements in the InAs/GaSb system obtained using
six-terminal small Hall devices with an edge channel length of 2 $\mu$m. We examine
three samples with different InAs layer thicknesses and correspondingly
different degrees of band overlap. In addition to the standard longitudinal
resistance, we examine non-local resistances for various current/voltage
contact configurations. For the optimal InAs layer thickness, the non-local
measurements reveal that, over a range of gate voltages, edge channel transport
is dominant with negligible bulk contribution. Due to the absence of bulk
transport, the non-local measurements allow us to extract the resistances of
individual edge channels, revealing reproducible fluctuations indicative of
inelastic scattering processes. Our results demonstrate that, by appropriately
designing the layer structure, the InAs/GaSb system can be tailored to have
conducting edge channels while maintaining a gap in the bulk region,
which is characteristic of a 2D TI band structure.
Our method is applicable not only to
the InAs/GaSb system but also to other 2D TI systems including the HgTe/HgCdTe
system, allowing us to study the properties of edge channel transport \cite{scattering-1,scattering-2,scattering-3,scattering-4,scattering-5}
even when the inelastic mean free path is shorter than the edge channel length and
quantized transport is absent.

\section{SAMPLES}

\begin{figure}[t]
\includegraphics[width=0.9\linewidth,clip]{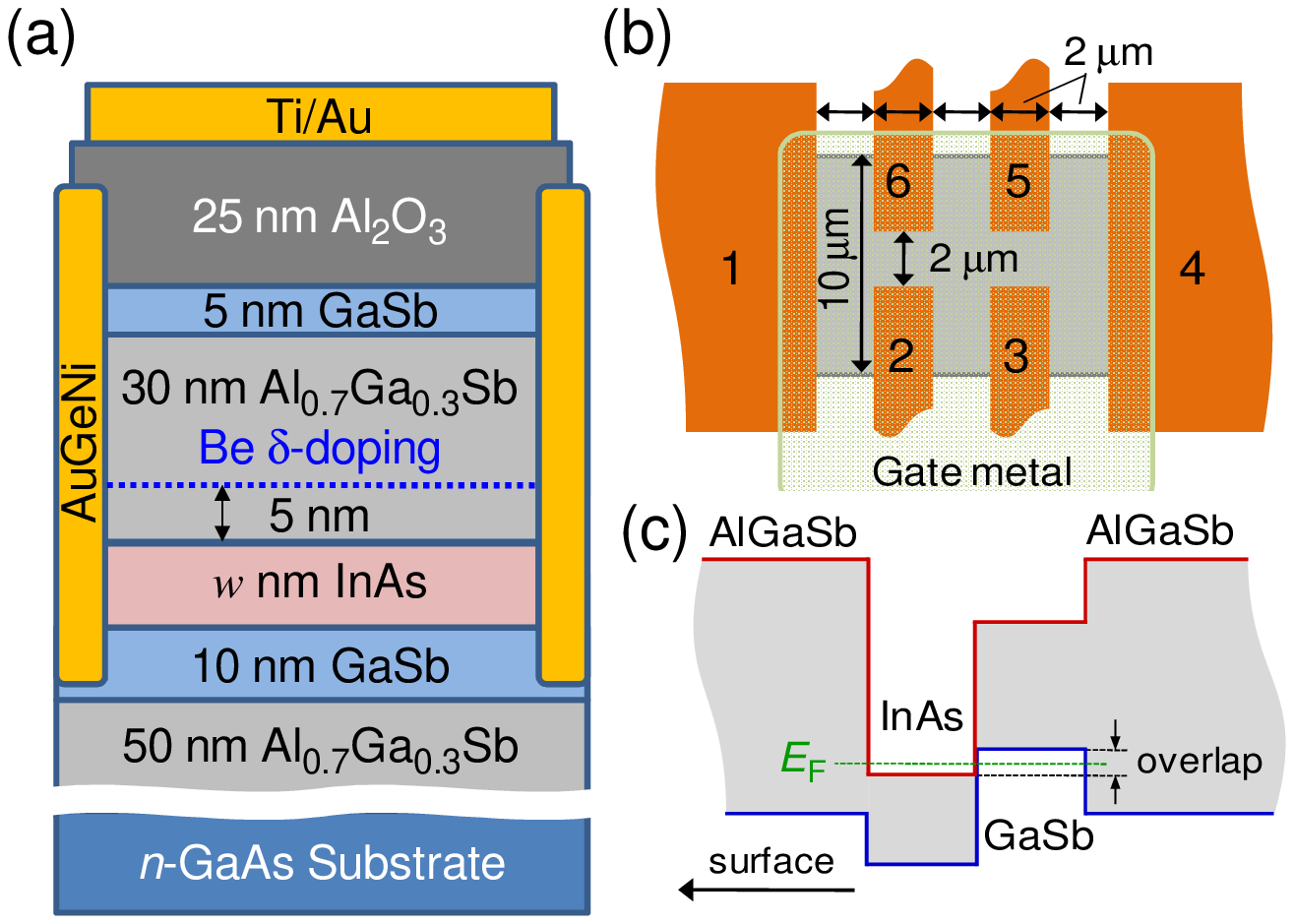}
\caption{(Color online) 
Schematic illustrations of the samples: (a)
Cross-sectional view of layer structure, (b) top view of
Hall bar, and (c) band profile along the growth direction.
Three samples with different InAs layer thicknesses
($w$ = 10, 12, and 14 nm) were examined.}
\label{fig1}
\end{figure}

The layer structure of the samples is shown schematically in Fig.~1(a).
InAs/GaSb heterostructures are grown by molecular beam epiatxy 
on an $n^{+}$-GaAs substrate, which serves as a back gate, with GaAs/AlAs and GaSb/AlSb
superlattice insulating buffer layers. We examined three heterostructures with
different InAs layer thicknesses $w$ (= 10, 12, 14 nm). 
To bring the Fermi level close to the energy gap region, the upper Al$_{0.7}$Ga$_{0.3}$Sb
barrier layer is delta-doped with [Be] $=5\times10^{11}$ cm$^{-2}$ at a
setback of 5 nm. \cite{Doping} 
Small Hall bar patterns as depicted in Fig.
1(b) are formed by using deep ultraviolet lithography and wet etching. The
Hall bars have six ohmic contacts formed by evaporating AuGeNi without
annealing. A Ti/Au top gate is evaporated on a 25-nm thick Al$_{2}$O$_{3}$
gate insulating layer deposited by atomic layer deposition. \cite{APEX} 
The gate metal entirely covers the active region of the Hall bar, including the
portion of each ohmic contact that overlaps the mesa. 
Ungated regions, if present, would remain $p$-type, which would then form $p$-$n$ junctions 
at the boundaries with gated regions when the latter are $n$-type. 
As a result, when the sample has a relatively large band
gap, a gated region would be electrically isolated from the ohmic contacts.
To avoid such situations, the gate metal was designed to entirely cover the
active region of the Hall bar, including the boundaries with the ohmic contacts.
The Hall bars are 10 $\mu$m wide and the distance between adjacent ohmic
contacts is 2 $\mu$m. Transport measurements are performed at $T=0.25$-4.3 K
using a lock-in technique with an excitation current of 1 nA and frequency of
13 Hz. All measurements are performed in a zero magnetic field and with a zero
back-gate voltage.

\section{EFFECTS OF I\lowercase{n}A\lowercase{s} LAYER THICKNESS}

\begin{figure}[t]
\includegraphics[width=0.9\linewidth,clip]{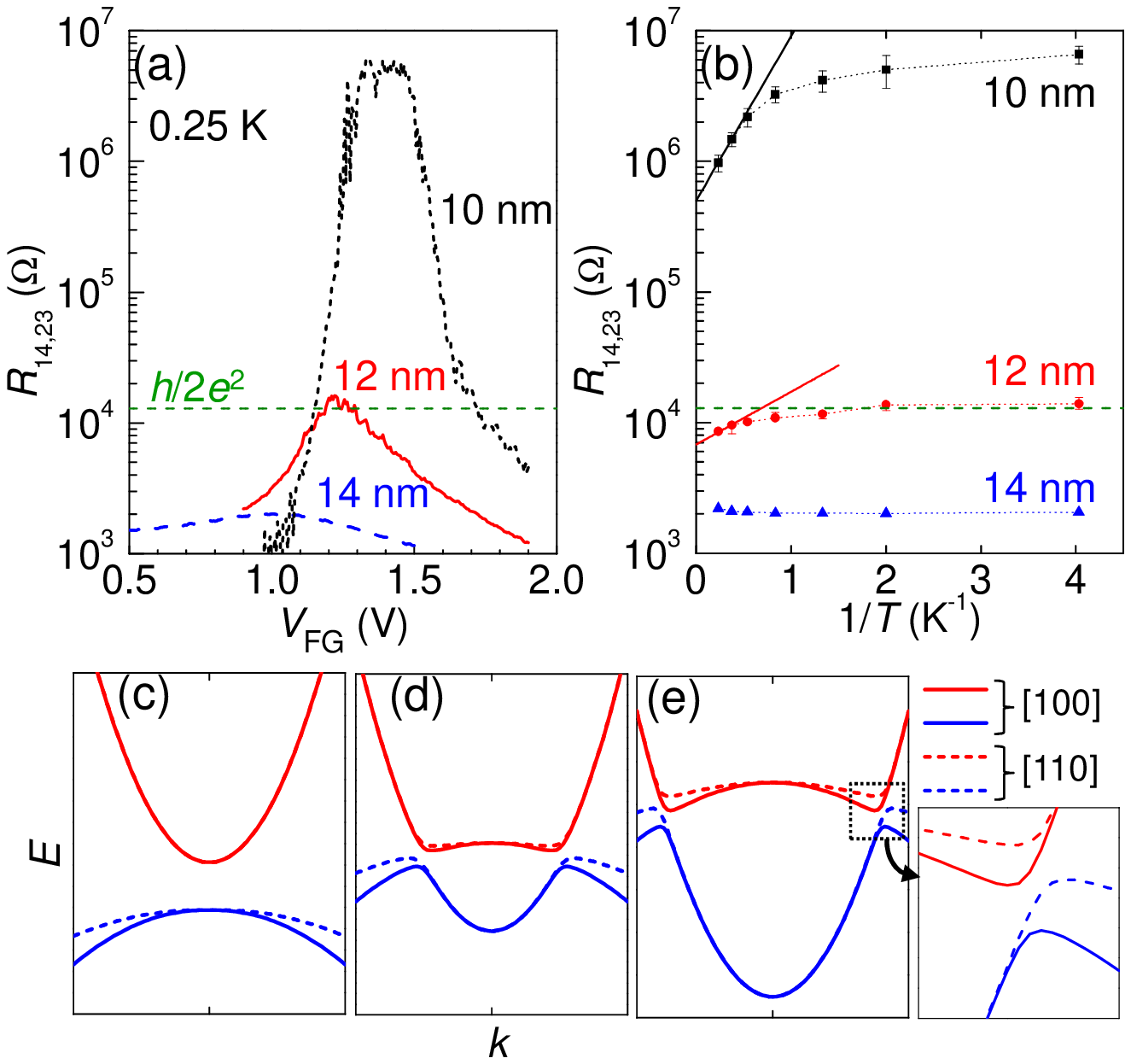}
\caption{(Color online) 
(a) Longitudinal resistance $R_{14,23}$ ($= V_{23}/I_{14}$)
at 0.25 K as a function of $V_\text{FG}$ for samples with different InAs thicknesses $w$. 
(b) Temperature dependence of $R_{xx}$ peaks plotted versus inverse temperature (1/$T$). 
Solid lines indicate fits to the high-temperature section of the data, 
which provide a crude estimate of the gap, ${\it \Delta} = 0.5$ and 0.2 meV, for $w = 10$ and 12 nm, respectively.  
(c)-(e) Schematic illustrations of energy band diagrams. 
The band overlap increases from (c) to (e). 
In (e), the region near the anti-crossing is shown in an enlarged view (right).
}
\label{fig2}
\end{figure}

We first show the impact of InAs layer thickness on the transport properties.
Figure 2(a) shows longitudinal resistance $R_{xx}$ (=$R_{14,23}$%
=$V_{23}/I_{14}$) as a function of front gate voltage ($V_{\text{FG}}$)
for the three samples with different $w$ values measured at 0.25 K. The suffixes
$R_{ij,kl}$ (=$V_{kl}/I_{ij}$) indicate the contacts used for driving the 
current ($i$, $j$) and measuring the resultant voltage ($k$, $l$). 
A maximum is observed in the $R_{xx}$ trace for all samples,
but the peak values differ significantly.
The peak resistance varies with $w$ by more than three
orders of magnitude, ranging from 2.0 k$\Omega$ for $w=14$~nm up to 8
M$\Omega$ for $w=10$ nm. The $w=12$ nm sample shows a peak resistance of
$\sim13$~k$\Omega$, which is more than two orders of magnitude smaller than
that of the $w=10$ nm sample. Hall measurements at $B$ = 1 T (data not shown)
confirm that, in all the samples, the majority carrier type changes from
$p$-type (holes) to $n$-type (electrons) as $V_{\text{FG}}$ is increased
across the $R_{xx}$ maximum. In those $V_{\text{FG}}$ ranges away from the
$R_{xx}$ maximum, the $R_{xx}$ value does not depend significantly on $w$.

The difference in the behavior of these three samples is corroborated by
examining the temperature dependence of $R_{xx}$ at its maximum [Fig.~2(b)].
The strongly insulating behavior of the $w = 10$ nm sample, with $R_{xx}$
increasing to several M$\Omega$ with decreasing temperature, clearly shows the
existence of a band gap. Fitting the high-temperature section of the data with
$R_{xx}\propto\text{exp}(\Delta/2k_\text{B}T)$ ($k_\text{B}$: Boltzmann
constant) provides a crude estimate of the energy gap, $\Delta=0.5$ meV. On
the other hand, the $w = 14$  nm sample shows almost no temperature dependence,
indicating a gapless band structure. In contrast, the $w = 12$ nm sample exhibits 
dual behavior; while the thermally activated behavior in the high temperature
regime suggests the existence of a band gap, at low temperatures $R_{xx}$ is
saturated and the system remains highly conductive. We note that the peak resistance
$\sim13$~k$\Omega$ of the $w = 12$ nm sample is close to the quantized value
$h/2e^{2}$ expected for the QSHE in the present contact geometry (where $h$ is
Planck's constant; $e$ is the elementary charge). 
Actually, the peak resistance for $w = 12$ nm varies 
from 12 to 50 k$\Omega$ among samples fabricated from the same wafer.
Even then, by comparison with $w = 10$ nm samples, the value is much lower and closer to
$h/2e^{2}$.

These results can be qualitatively understood from the band diagrams shown 
schematically in Fig.~1(c) and Figs.~2(c)-2(e). When the InAs layer is
thin, due to the strong confinement of the electron wavefunction, the electron
subband in InAs lies above the hole subband in GaSb, resulting in a normal
semiconductor-like band structure with a band gap [Fig.~2(c)]. As the InAs
layer is made thicker, the electron subband becomes lower in energy and, at
some point band inversion occurs with the hole subband in GaSb, where band
hybridization across the InAs/GaSb interface opens a small gap at the band
crossing points [Fig.~2(d)]. 
Spin-orbit interaction induces linearly dispersing bands inside
this gap and they connect the conduction and valence bands, which
correspond to the helical edge channels of a 2D TI. \cite{InAs_GaSb} 
When the Fermi level is in this gap, transport is governed by the helical edge
channels at low temperatures. As the InAs layer thickness is increased
further, the band overlap increases; however, the influence of the GaSb valence
band anisotropy becomes more important because the anti-crossings occur at
larger wave numbers \textbf{k}. \cite{Du} 
While the gap remains open for each
\textbf{k} direction, now the energy position of the gap strongly depends on
the \textbf{k} direction [Fig.~2(e)]. When the anisotropy of the gap position
exceeds the size of the gap, the upper and lower bands along different
\textbf{k} directions overlap in energy, leading to a semi-metallic band
structure without a band gap.

\section{NON-LOCAL MEASUREMENTS}

\begin{figure}[bhtp]
\includegraphics[width=0.9\linewidth,clip]{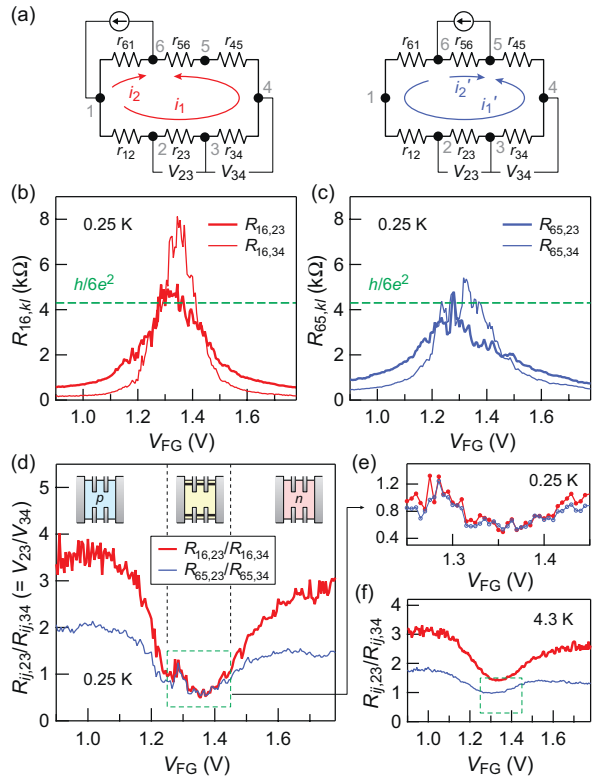}
\caption{(Color online) 
(a) Effective circuit model describing edge transport in the topological insulating phase.
Left and right panels represent current/voltage contact configurations for the non-local measurements 
in (b) and (c), respectively.  
See main text for details. 
(b) and (c) non-local resistances of the $w = 12$ nm sample at 0.25 K as a function of $V_\text{FG}$: 
(b) $R_{16,23}$ and $R_{16,34}$ and (c) $R_{65,23}$ and $R_{65,34}$. 
(d) Ratio between non-local resistances measured
with adjacent voltage-probe pairs (2-3 and 3-4) at 0.25 K. 
The two traces show results for different current paths (1-6 and 6-5). 
(e) Magnified view of the data in the dashed rectangle in (d). 
(f) Similar measurements to (d) at 4.3 K.
}
\label{fig3}
\end{figure}

To clarify the transport mechanism in the $w =12$ nm sample, we investigate
non-local resistances. In a topologically insulating phase, the bulk is
insulating and the current would flow only through the helical edge channels
that form along the sample edges. The current and voltage distributions in
such a topological phase can be understood using the effective circuit model
as shown in Fig.~3(a), in which the system can be described as a network of
channel resistances $r_{j,j+1}$ ($j=1$,$2$,...; mod $6$). The injected current
is partitioned into two paths along the sample edges connecting the source and
drain contacts in the clockwise ($i_{2}$) and counterclockwise ($i_{1}$)
directions. In an ideal QSHE system without inelastic scattering, each
channel has a quantized conductance of $e^{2}/h$ for counterpropagating
opposite-spin components, so that the current, when driven through adjacent
contacts, is partitioned at a ratio of 1:5 in a six-terminal device. Thus, a
quantized resistance of $h/6e^{2}$ should be observed for the all
configurations in an ideal system without inelastic scattering.

\begin{figure*}[phbt]
\includegraphics[width=0.8\linewidth,clip]{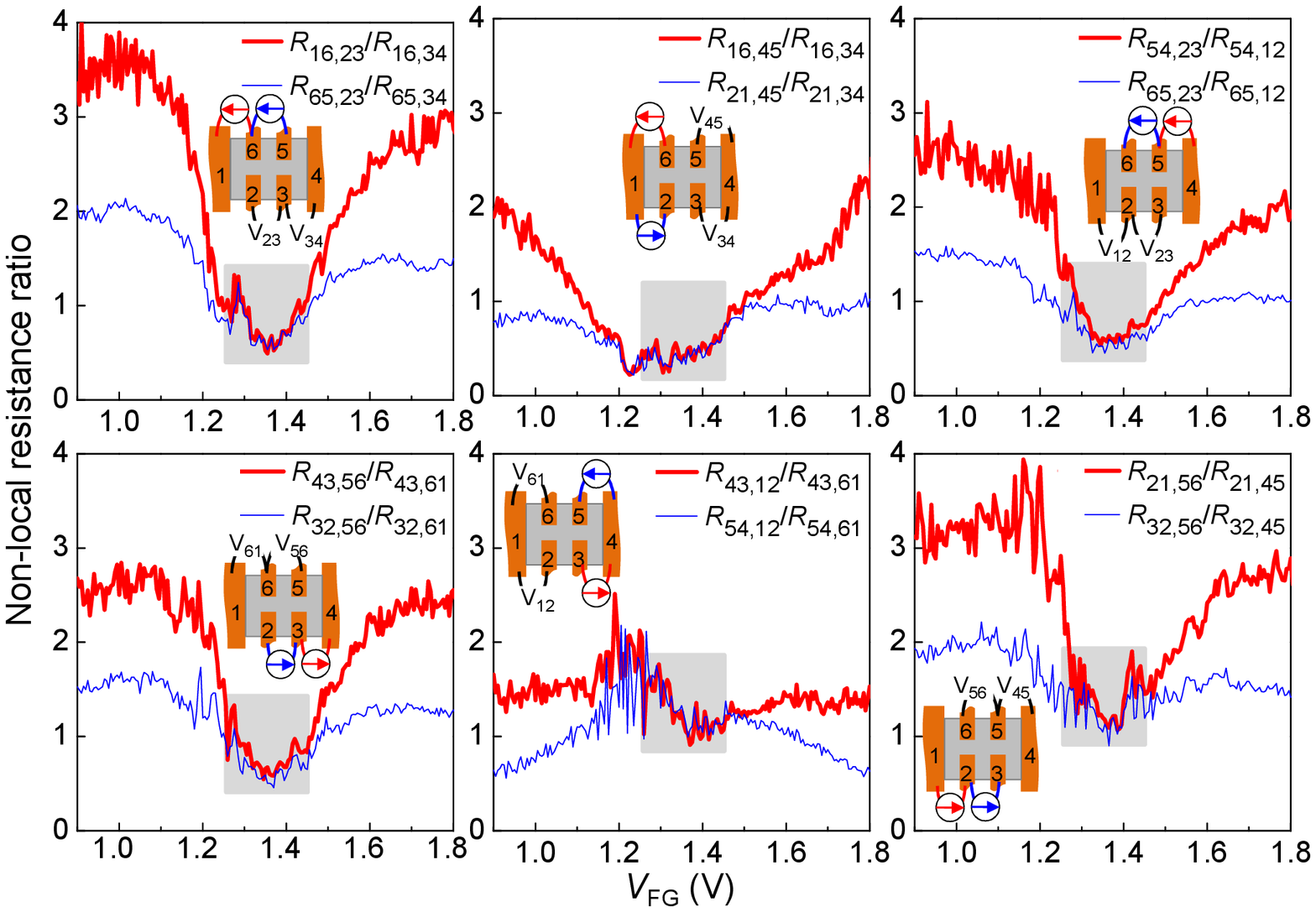}
\caption{(Color online) 
Results of non-local measurements on the $w$ = 12 nm sample for various current/voltage contact configurations at 0.25 K. 
Each panel corresponds to a given set of adjacent voltage-probe pairs shown in the inset. 
Ratio between non-local resistances measured with these voltage-probe pairs are plotted as a function of $V_\text{FG}$. 
The two traces in each panel represent measurements for the two different current injection/ejection paths shown in the inset. 
Note that each measurement uses only one of the two current sources shown in the inset. 
The gray background represents the region where the Fermi level is expected to be in the gap.
}
\label{fig4}
\end{figure*}

Figure 3(b) shows the non-local resistances $R_{16,23}$ and $R_{16,34}$
measured at 0.25 K as a function of $V_\text{FG}$. These two measurements
use the same current injection/ejection path (1-6), 
and differ only in the contact pair (2-3 or 3-4) used for measuring the voltage. 
Figure 3(c) shows similar measurements
($R_{65,23}$ and $R_{65,34}$) for annother current injection/ejection path (6-5). 
In the range $1.25<V_\text{FG}\ <1.45$ V, where the Fermi level is expected to be
in the bulk gap, all the measured non-local resistances show a tendency to
approach $h/6e^{2}$ as expected for the QSHE (indicated by dashed line).
However, there are significant deviations from $h/6e^{2}$. 
Intriguingly, the non-local resistances exhibit sharp fluctuations as a function of
$V_\text{FG}$. These resistance fluctuations are reproducible with
amplitudes larger than the background noise level (see Appendix).

As we show below, focusing on the ratio between non-local resistances measured
for adjacent voltage-probe pairs reveals that transport is dominated by edge
channels despite the significant fluctuations and deviation from the quantized
value. In Fig.~3(d), we plot resistance ratios $R_{16,23}/R_{16,34}$ and
$R_{65,23}/R_{65,34}$ calculated from the data in Figs.~3(b) and 3(c) as a
function of $V_\text{FG}$. Physically, these ratios represent the voltage
ratio $V_{23}/V_{34}$ that appears for a given current injection/ejection
path, 1-6 or 6-5. If bulk transport is dominant, the voltage ratio
$V_{23}/V_{34}$ would simply reflect the current distribution inside the bulk,
which is determined solely by the current injection/ejection path and does not
depend on sample conductivity. Therefore, the ratios should take different
values for different current paths. This is actually seen in the
$V_\text{FG}$ ranges where the Fermi level is outside the gap; at
$V_\text{FG}<1.25$ V ($p$-type region) and $V_\text{FG}>1.45$ V ($n$-type
region), $R_{16,23}$/$R_{16,34}$ and $R_{65,23}$/$R_{65,34}$ do not coincide
with each other.

In contrast, if edge transport is dominant as in Fig.~3(a), the voltage ratio
$V_{23}/V_{34}$ should directly reflect the relative magnitude of the
resistances $r_{23}$ and $r_{34}$ of the relevant edge channels, since the
same current ($i_{1}$ or $i_{1}^{\prime}$) flows through them for a given
current injection/ejection path [Fig.~3(a)]. An important corollary of the
edge transport is that the ratio $V_{23}/V_{34}=r_{23}/r_{34}$ is an intrinsic
property of the sample and therefore does not depend on the current
injection/ejection path. Note that this holds for arbitrary values of
$r_{j,j+1}$, that is, even if they deviate from the quantized value and
consequently the current partition ratio deviates from 1:5, provided that the
bulk transport is negligible. Indeed, we observe that in the gap region
($1.25$ $<V_\text{FG}<1.45$~V) the resistance ratios for the two current
paths coincide with each other [Fig.~3(d)], including the fine details of the
fluctuations [Fig.~3(e)]. 

Figure~4 compiles the results of similar non-local measurements on the same sample for all possible contact configurations. 
In each panel, the ratio between the non-local resistances measured 
with a given set of adjacent voltage-probe pairs is plotted as a function of $V_\text{FG}$. 
The two traces in each panel represent the results obtained for the two different current injection/ejection paths shown in the inset. 
The data demonstrate that, for all configurations, the resistance ratios become independent of the current path in the gap region. 
These results constitute compelling evidence that the transport in the gap region is dominated by edge channels.

Figure~3(f) shows the resistance ratios measured at 4.3 K for the same contact
geometries as in Fig.~3(d) at 0.25 K. The behavior outside the gap region ($V_\text{FG} 
<1.25$ V and $V_\text{FG}>1.45$ V) is similar to that at 0.25 K, as
expected for the bulk conduction. On the other hand, in the gap region, the
resistance ratios for the two current paths no longer coincide with each
other. This indicates that at high temperatures there is a finite current that
flows directly between non-adjacent contacts through the bulk region. This is
reasonable because at 4.3 K thermally activated carriers contribute to the
bulk conduction; it is also consistent with the temperature dependence of
$R_{xx}$ [Fig.~2(b)].

By comparing a series of InAs/GaSb devices with different
lengths and widths, Knez and co-workers were able to separate out the
contributions of bulk and edge transport; they found that there was a significant
contribution from the bulk conductivity $g_{\text{bulk}}\approx5e^{2}/h$
at 0.3 K in their samples. \cite{Du} 
Assuming the same bulk
conductivity for our sample geometry would yield $R_{14,23}=1.5$ k$\Omega$,
which is much smaller than the measured value for the $w=12$ nm sample. This
clearly demonstrates that, by appropriately designing the layer structure, the
InAs/GaSb system can be tailored to have conducting edge channels while
keeping a gap in the bulk region, as predicted in Ref. \onlinecite{InAs_GaSb}.

Owing to the fact that bulk conduction in our sample is negligible at low
temperatures, we are able to determine the resistance ratios for all the
adjacent contact pairs by performing non-local measurements while sequentially
changing the current/voltage contact configurations. This allows us to extract
the individual resistances of all the edge channels as a function of
$V_{\text{FG}}$ [Fig.~5(a)]. The channel resistances deduced in this way
range from 13 to 70~k$\Omega$, which deviate greatly from the expected quantized
value, $h/e^{2}\approx25.8$~k$\Omega$. To verify the consistency of
measurements, the longitudinal resistance $R_{14,65}$ is reconstructed from
these channel resistances using the circuit model in Fig.~3(a). Despite the
large scattering of the individual channel resistances and experimental errors
involved, the reconstructed $R_{14,65}$ agrees well with measured value [Fig.~5(b)].

\begin{figure}[t]
\includegraphics[width=0.95\linewidth,clip]{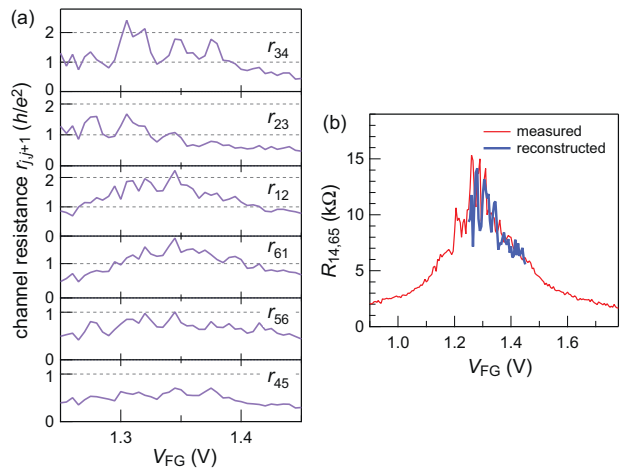}
\caption{(Color online) 
(a) Resistances of individual edge channels deduced from non-local resistance measurements at 0.25 K. 
(b) Comparison of longitudinal resistance $R_{14,65}$ reconstructed from the individual channel resistances
shown in (a) with that measured directly.
}
\label{fig5}
\end{figure}

As the data in Fig.~5(a) demonstrate, non-local measurements allow us
to examine the scattering processes occurring in a single edge channel. The
resistances exhibit fluctuations indicative of energy-dependent inelastic
scattering. 
The amplitude of the fluctuations ranges up to about $h/e^{2}$. 
Inelastic scattering can occur when there are electron or hole puddles
arising from spatial potential fluctuations near the helical edge
channels that serve as reservoirs in the same way as ohmic contacts do. 
Hence these puddles increase the resistance in the same way 
as additional ohmic contacts embedded along the edge channels. \cite{supporting,Glazman} 
It is possible that the observed $V_{\text{FG}}%
$-dependent fluctuations represent the charging effects of localized puddles that
act as scatterers by exchanging electrons with the edge channels.
We should note that these scatterers always act to increase the channel
resistance. On the other hand, if there is a large puddle 
that is capable of exchanging carriers with both of two adjacent contacts, it will act to decrease
the channel resistance; this might explain the result where $r_{45}$ and
$r_{56}$ are smaller than $h/e^{2}$. Figure 5(a) also shows that the $V_{\text{FG}}$ range 
that satisfies $r_{j,j+1}\geq h/e^{2}$ varies slightly with
the position of the edge, suggesting that long-range sample inhomogeneity
makes it difficult to tune the Fermi level in the gap region over the entire
sample area. Further investigations are necessary to clarify the exact
scattering mechanism.

\section{SUMMARY}

In summary, we have investigated transport in InAs/GaSb heterostructures with
different InAs layer thicknesses. From systematic non-local resistance
measurements, we have shown that a sample with 12-nm-thick InAs layer has a bulk
gap and edge channel transport. Our method offers a way of elucidating the
properties of 2D TIs even when the inelastic mean free path is shorter than
the edge channel length and quantized transport is absent.

\newpage

\acknowledgments
We thank Yoshito Ishikawa, Satoshi Sasaki, and Mineo Ueki for their help with the experiments.

\appendix*
\section{}

The resistance fluctuation observed for the $w$ = 12 nm sample reflects the change in gate voltage. 
This can be verified by comparing it with the time-dependent fluctuation of the signal. 
Figure 6(a) shows the non-local resistance $R_{32,45}$ as a function of  $V_\text{FG}$, 
where the resistance was monitored at each  $V_\text{FG}$ for 60 sec with a 1-sec interval. 
The trace represents the average taken at each  $V_\text{FG}$ 
and the vertical bars indicate the amplitude of the time-dependent fluctuation at each  $V_\text{FG}$. 
The dashed rectangular region in Fig. 6(a) is enlarged in Fig. 6(b). 
It can be seen that the amplitude of the time-dependent fluctuation including background noise 
is much smaller than that of the  $V_\text{FG}$-dependent resistance fluctuations.

~
\begin{figure}[hpbt]
\includegraphics[width=0.9\linewidth,clip]{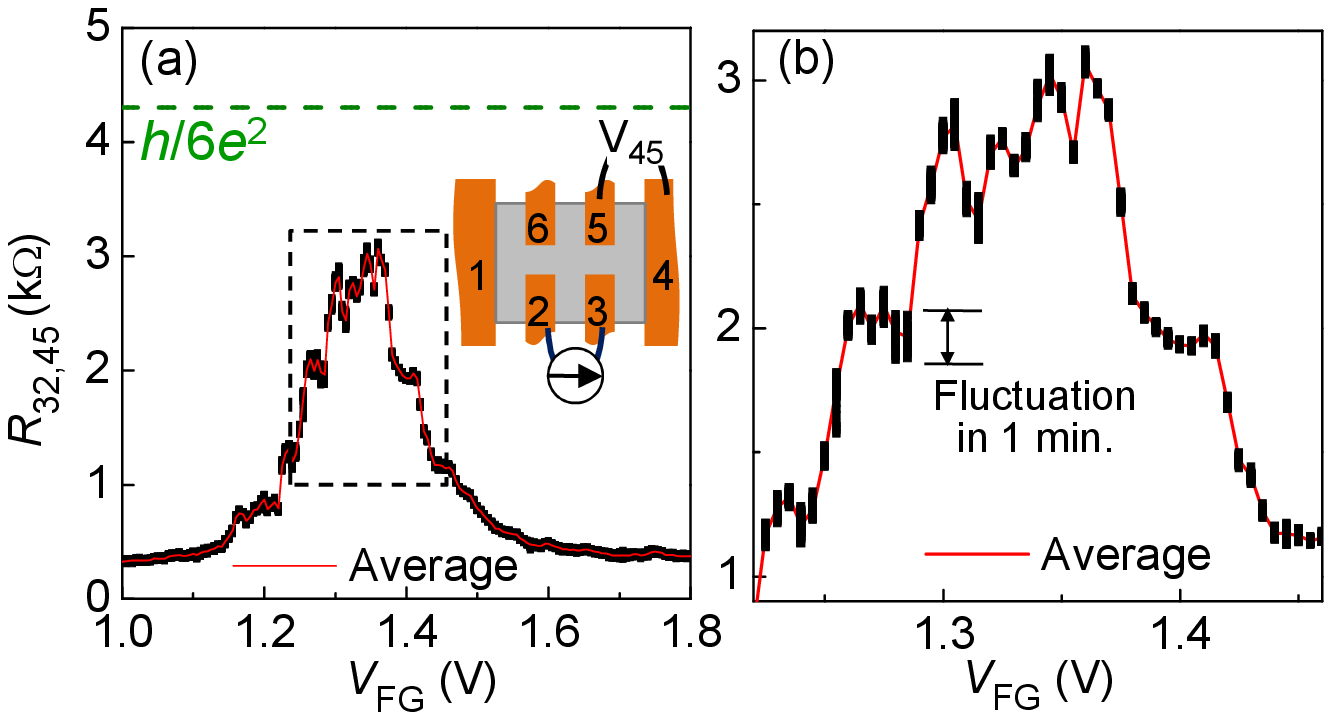}
\caption{(Color online) 
(a) Non-local resistance $R_{32,45}$ as a function of $V_\text{FG}$ for the $w$ = 12 nm sample ($T$ = 0.25 K). 
The trace represents the average taken over 1 min. at each $V_\text{FG}$ 
and the vertical bars indicate the amplitude of the time-dependent fluctuation at each $V_\text{FG}$. 
Inset shows the measurement configuration. (b) Magnified view of the data shown in the dashed rectangle in (a).
}
\label{fig6}
\end{figure}

\end{document}